\documentclass[aps,pre,superscriptaddress,nofootinbib,twocolumn,10pt]{revtex4-2}

\usepackage[T1]{fontenc}
\usepackage[utf8]{inputenc}
\usepackage{amsmath,amssymb,amsthm}
\usepackage{graphicx}
\usepackage{xcolor}
\usepackage{hyperref}
\usepackage{float}
\makeatletter
\let\newfloat\newfloat@ltx
\makeatother
\usepackage{algorithm}
\usepackage{algpseudocode} 

\usepackage{ifthen}

\newcommand{\new}[1]{\textcolor{black}{#1}}

\begin{document}

\title{Demonstrating Real Advantage of Machine-Learning-Enhanced Monte Carlo for Combinatorial Optimization}

\author{Luca Maria Del Bono}
\affiliation{Dipartimento di Fisica, Sapienza Università di Roma, Piazzale Aldo Moro 5, Rome 00185, Italy}

\author{Federico Ricci-Tersenghi}
\email{federico.ricci@uniroma1.it}
\affiliation{Dipartimento di Fisica, Sapienza Università di Roma, Piazzale Aldo Moro 5, Rome 00185, Italy}
\affiliation{CNR-Nanotec, Rome unit, Piazzale Aldo Moro 5, Rome 00185, Italy}
\affiliation{INFN, sezione di Roma1, Piazzale Aldo Moro 5, Rome 00185, Italy}

\author{Francesco Zamponi}
\email{francesco.zamponi@uniroma1.it}
\affiliation{Dipartimento di Fisica, Sapienza Università di Roma, Piazzale Aldo Moro 5, Rome 00185, Italy}

\begin{abstract}
Combinatorial optimization problems are central to both practical applications and the development of optimization methods.
While classical and quantum algorithms have been refined over decades, machine learning--assisted approaches are comparatively recent and have not yet consistently outperformed simple, state-of-the-art classical methods.
Here, we focus on a class of Quadratic Unconstrained Binary Optimization (QUBO) problems, specifically the challenge of finding minimum energy configurations in three-dimensional Ising spin glasses. We use a Global Annealing Monte Carlo algorithm that integrates standard local moves with global moves proposed via machine learning. We show that local moves play a crucial role in achieving optimal performance. Benchmarking against Simulated Annealing and Population Annealing, we demonstrate that Global Annealing not only surpasses the performance of Simulated Annealing but also exhibits greater robustness than Population Annealing, maintaining effectiveness across problem hardness and system size without hyperparameter tuning. These results provide clear and robust evidence that a machine learning--assisted optimization method can exceed the capabilities of classical state-of-the-art techniques in a combinatorial optimization setting.
\end{abstract}

\maketitle

Combinatorial optimization problems appear in a variety of real-world applications as well as in fundamental theoretical studies.
They consist of finding an optimal state, specified by $N$ discrete variables, that minimizes an objective function over a finite, but exponentially large in $N$, set. They include, but are not limited to, maximum-satisfiability (MAX-SAT)~\cite{li2009maxsat}, graph coloring~\cite{jensen1995graph}, MAX-CUT~\cite{goemans1995improved}, maximum independent set~\cite{garey1979computers}, job scheduling~\cite{pinedo2016scheduling}, set cover~\cite{johnson1974approximation, chvatal1979setcover}, and the traveling salesman problem~\cite{lawler1985tsp}.

Many different algorithms have been introduced in the course of the years to study combinatorial optimization. Exact deterministic solvers are available, but their applicability is limited to moderately large sizes \cite{rendl2007maxcut, krislock2017biqcrunch, gusmeroli2022biqbin, CJMM22,gurobi, nakhle2025gta} due to an exponential increase in the computational cost.
Hence, state-of-the-art scalable solvers for combinatorial optimization are instead based on simple {\it local} stochastic rules, in which one or a few variables are updated at each step. Starting from a random assignment of variables, local moves are proposed by some heuristics, and accepted according to the achieved gain in the objective function. A prominent example is Simulated Annealing (SA)~\cite{kirkpatrick1983sa}, but other effective heuristics exist \cite{marinari2000effects, boettcher2004extremal, alava2008circumspect, benlic2013breakout, angelini2019monte, bernaschi2021we}. 
Quantum algorithms such as Quantum Annealing \cite{kadowaki1998qa, farhi2001aqc, santoro2002theory, das2008rmp, fioroni2025entanglement} and Quantum Approximate Optimization Algorithms \cite{farhi2014qaoa, brady2024iterative} have also been proposed~\cite{farhi2020needs}. However, these algorithms often fail to find the best solution and can only find approximate ones, often with large relative errors. To date, they do not seem to outperform classical ones~\cite{bapst2013quantum}.

Recently, it has been proposed to use machine-learning~(ML) methods as a way to generate better heuristics for stochastic search algorithms, in particular because such methods can propose {\it global} moves in which most of the variables are updated in a single step~\cite{bengio2021mlco, dai2017learnco, kool2019attention, pahng2020predicting, korol2022calculation, fan2023searching, feng2025sequential, krylova2025unsupervised}. 

Yet, until now, most results have been limited to moderately large sizes and do not seem to outperform classical algorithms. 
Claims of superiority~\cite{schuetz2022combinatorial} have been contested \cite{angelini2023modern, boettcher2023inability}. Some works have focused in particular on the task of finding the minimum energy configuration of spin glass models~\cite{sanokowski2022one, dobrynin2025nonlocal, korol2022calculation, sanokowski2025scalable}. Also in this case, claims of superiority~\cite{fan2023searching} have been contested~\cite{boettcher2023deep, fan2023reply}. 
Therefore, it is still unclear whether a novel ML-based method can outperform or even perform comparably with state-of-the-art classical algorithms.

To give a clear answer to this question, in this work, we consider a hard benchmark for combinatorial optimization, namely the three-dimensional Edwards-Anderson spin glass model that provides instances of Quadratic Unconstrained Binary Optimization (QUBO). We perform a fair comparison of an ML-assisted Global Annealing (GA) algorithm with two state-of-the-art solvers, namely SA and Population Annealing (PA). 
We present robust evidence that an ML-assisted algorithm can outperform state-of-the-art solvers under controlled conditions.

More specifically, our main results are as follows.
\begin{itemize}

    \item We show (Fig.~\ref{fig:success_rates_twoinstances}) that the GA procedure requires a combination of ML-assisted global moves and simple local moves to be effective, thus confirming intuition from previous theoretical works \cite{gabrie2022adaptive, del2025performance}.
    
    \item We compare the runtimes of SA, PA, and GA, which can all be implemented in a similar way using the \texttt{torch} \cite{NEURIPS2019_9015} environment, thus allowing for a fair comparison of wall-clock times. 
    We show that, for large systems of $N = 10^3$ variables, GA consistently outperforms SA (Fig.~\ref{fig:success_rates_twoinstances}). Moreover, while it performs worse than PA on easy instances of the problem, it appears to perform on par or better on harder instances, thus showing improved robustness with respect to instance-to-instance hardness fluctuations (Figs.~\ref{fig:success_vs_time_L10} and \ref{fig:scatter_plot_times}).
    
    \item Once scaled to even larger systems of $N = 14^3 = 2744$ spins, we show that GA consistently outperforms PA (Fig.~\ref{fig:success_vs_time_L14}). We note that this is achieved by using the same hyperparameters as in the $N = 10^3$ case, hence showing greater robustness of GA to changes in problem specification.
    We also point out that these sizes approach the limit for what can be solved with state-of-the-art algorithms. 
    
\end{itemize}
We stress once again that our comparison is performed using one of the hardest QUBO benchmarks, with problems of large size, under comparable implementations, and using wall-clock times. Hence, we believe that our analysis is fair, and we can make a robust claim of the superiority of ML-assisted algorithms. 
We also provide some tentative explanations for why GA outperforms state-of-the-art solvers by examining how the combinatorial space is explored along the annealing process~(Fig.~\ref{fig:Pq}).

The paper is organized as follows. In the Algorithms section, we define precisely the optimization problem under study and the algorithms we consider.
In the Results section, we present the main numerical results for our study. In particular, we present extensive results for $N = 10^3$ spins and additional results for $N = 14^3$ spins. In the Discussion section, we discuss our results and present possible future directions for research. Finally, in the Materials and Methods section, we give a detailed description of the algorithms used and of the details of our simulations. Additional data can be found in the Supporting Information.

\section{Algorithms}
\label{sec:algorithms}

\subsection{Combinatorial optimization}

A wide range of combinatorial optimization problems belongs to the QUBO class~\cite{boettcher2019}, where one wants to find $x^* = \text{argmin}_x \; E(x)$ for
\begin{equation}
    E(x) =  - \sum_{i < j} Q_{ij} \, x_i x_j - \sum_{i} b_i x_i \ ,
\end{equation}
where $x=(x_1,\cdots,x_N)$, with $x_i \in \{0, 1\}$, being a set of $N$ boolean variables.
In statistical physics language, the QUBO class is equivalent to Ising optimization~\cite{lucas2014ising}, where one wants to find a \textit{configuration} of $N$ binary Ising variables, $\sigma = (\sigma_1, \dots, \sigma_N)$ with $\sigma_i \in \{ -1, +1\}$, called \textit{spins}, that minimizes an energy function
\begin{equation}\label{eq:genericH}
H(\sigma) = - \sum_{i < j} J_{ij} \, \sigma_i \sigma_j - \sum_{i} h_i \sigma_i \ ,
\end{equation}
hence
\begin{equation}\label{eq:argminH}
    \sigma^* = \text{argmin}_\sigma \; H(\sigma) \ .
\end{equation}
We note that, in general, there can be multiple degenerate minima, i.e., solutions of ~\eqref{eq:argminH}. 
Many of the previously cited optimization problems can be recast as Ising optimization problems \cite{lucas2014ising}. 

The energy function, ~\eqref{eq:genericH}, also plays a very important role in statistical physics, where it is associated with the Gibbs-Boltzmann (GB) distribution,
\begin{equation}\label{eq:GB}
    \rho_\text{\tiny GB}(\sigma) \propto e^{-\beta H(\sigma)} \ ,
\end{equation}
which defines the probability distribution of configurations at thermal equilibrium with temperature $T =\beta^{-1}$. Depending on the specific choice of the couplings $J_{ij}$ and the fields $h_i$, ~\eqref{eq:GB} can describe a variety of different systems and phenomena, such as the paramagnetic-ferromagnetic transition at the Curie temperature \cite{onsager1944crystal}.
A particularly interesting choice is considering random symmetric coupling $J_{ij} = J_{ji} \sim \mathcal{N}(0,1)$ between nearest neighbors on a $d$-dimensional square lattice. This choice defines the Edwards-Anderson (EA) model \cite{edwards1975theory} for \textit{spin glasses}, whose energy landscape is particularly complex and hard to optimize for $d \geq 3$. In particular, it has been shown that finding the minimum energy configuration of the system, i.e., solving ~\eqref{eq:argminH}, is NP-hard for any $d \geq 3$ \cite{barahona1982computational, fan2023searching}. In this paper, we are interested in studying the $d=3$ EA model without external magnetic fields ($h_i = 0$), as it provides a set of hard benchmarks in the QUBO class.

\begin{figure*}[t]
    \centering
    \includegraphics[width=\linewidth]{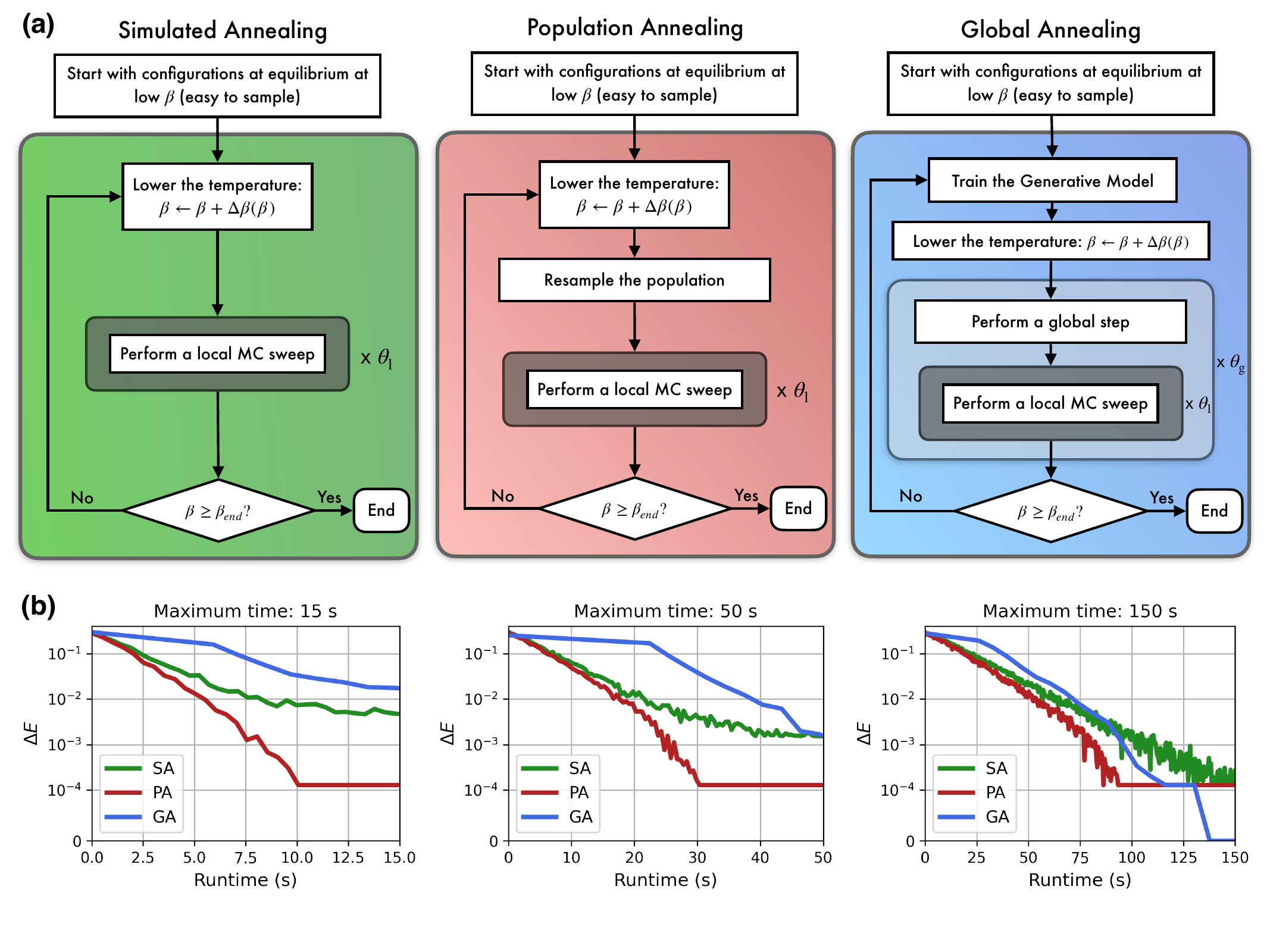}
    \caption{(\textit{a}) Schematic description of the three algorithms considered in this work: the well-known Simulated Annealing (SA) and Population Annealing (PA), and the novel machine-learning-assisted Global Annealing (GA). $\theta_l$ and $\theta_g$ represent the number of local and global steps, respectively. \new{At each annealing step, the temperature is updated according to $\beta \gets \beta + \Delta \beta(\beta)$, where $\Delta \beta(\beta)$ is a (generally not constant) temperature schedule.} (\textit{b}) Example of how the minimum energy decreases during three different runs of various lengths for a given instance of the problem. The difference $\Delta E$ between the minimum energy found by the algorithms and the exact one (here found using the Gurobi solver~\cite{gurobi}) is plotted as a function of the simulation time for a short (15 seconds), a medium (50 seconds), and a long (150 seconds) run. As the running time increases, the three algorithms manage to progressively find lower energy states. \new{The chosen instance is the same \textit{hard instance} used for the right panel of Fig.~\ref{fig:success_rates_twoinstances} and has $L = 10$, $N = 10^3$.}}
    \label{fig:algorithms}
\end{figure*}

\subsection{Sampling solvers}

Apart from the methods mentioned in the Introduction, energy optimization can also be performed using sampling algorithms, that is, techniques whose broader goal is to obtain configurations distributed according to ~\eqref{eq:GB}. Indeed, when $\beta\to\infty$, the probability distribution in ~\eqref{eq:GB} concentrates on the configuration(s) of minimum energy. Therefore, if one can sample correctly from $\rho_\text{\tiny GB}(\sigma)$ at all temperatures, then one can also find the minimum energy configuration of the system by slowly decreasing the temperature down to zero. This idea was first implemented by the famous SA algorithm~\cite{kirkpatrick1983sa}.
Following this idea, several improved sampling algorithms have been developed, such as Simulated Tempering~\cite{marinari1992simulated}, Parallel Tempering \cite{hukushima1996exchange} and its modifications \cite{houdayer2001cluster, zhu2015efficient}, and PA~\cite{hukushima2003population, machta2010pa,wang2015pa}. For completeness, we mention that other classical algorithms based on dynamical systems~\cite{goto2021high, ekanayake2025different}, or tensor-networks~\cite{smierzchalski2025spinglasspeps, chen2025batchtnmc} exist.

More recently, following the machine learning revolution, it was proposed to use generative modeling architectures to assist sampling~\cite{carleo2017solving}. This led to a variety of proposals, using architectures such as autoregressive networks~\cite{wu2019solving, wang2025enhancing} and transformers \cite{bunaiyan2025isingformer}, normalizing~\cite{noe2019boltzmann, invernizzi2022skipping, dibak2022temperature, jung2024normalizing} and equivariant~\cite{kohler2020equivariant, kanwar2020equivariant, albergo2019flow, gerdes2022learning, de2021scaling, schonle2025sampling} flows, diffusion models~\cite{biroli2023generative, bae2025diffusion, hunt2024accelerating, sanokowski2025scalable}, stochastic interpolants~\cite{albergo2023stochastic, chen2025scale}, Boltzmann machines~\cite{decelle2024restricted} and renormalization-group inspired models~\cite{marchand2022wavelet, masuki2025generative}, which were implemented in different algorithms such as Sequential Tempering~\cite{mcnaughton2020boosting}, Adaptive Monte Carlo~\cite{gabrie2022adaptive} and \new{variational} neural annealing \cite{hibat2021variational, inack2022neural}. Additional work has been carried out to use non-generative techniques ~\cite{galliano2024policy, tzivrailis2025uncertainty}, as well as mixing quantum and machine-learning methods~\cite{scriva2023accelerating}.

Despite the numerous and diverse ML-assisted algorithms proposed, there is currently no evidence that any of these can be effective in solving challenging optimization problems.
A proper benchmarking of a solver based on an ML-assisted sampling solver approach is the scope of this work.

\subsection{Details of the implemented algorithms}

In this work, we compare three different annealing techniques (that is, techniques in which the temperature is monotonously lowered): two classical algorithms, Simulated Annealing and Population Annealing, and the machine-learning-assisted Global Annealing.
The choice of SA and PA as benchmarks is motivated by two main reasons: {\it (i)} they are among the algorithms currently achieving the best results, and can therefore be regarded as state-of-the-art solvers for ~\eqref{eq:argminH}; and {\it (ii)} their implementation is very similar to that of GA, which ensures a fair comparison (Fig.~\ref{fig:algorithms}).
More specifically:
\begin{itemize}
    \item \textbf{Simulated Annealing (SA)}~\cite{kirkpatrick1983sa, caracciolo2023simulated} uses classical local Monte Carlo (MC) moves, in which a single variable $\sigma_i$ is updated at each time step, by flipping its sign~\cite{newman1999monte}. A set of $N$ such moves is referred to as a Monte Carlo Sweep (MCS). The algorithm attempts to sample according to the GB distribution, \eqref{eq:GB}, at different temperatures: the sampling starts at high temperature, and then $T$ is progressively lowered in small steps. Since \eqref{eq:GB} collapses on the minimum energy states when $\beta \to \infty$, as the temperature is lowered, the typical energy of the states sampled by the local MC  progressively decreases. If equilibration can be maintained at all temperatures, one obtains a solution of ~\eqref{eq:argminH}. Otherwise, one might get stuck at a higher energy, only achieving an approximate solution to the optimization problem.
    \item \textbf{Population Annealing (PA)}~\cite{hukushima2003population, machta2010pa,wang2015pa, amey2018analysis, jung2025numerical} works similarly to Simulated Annealing, but evolving a whole population of configurations at once. At each step, when the temperature is lowered, the population of configurations is resampled in order to better adapt to the lower temperature. Previous studies~\cite{wang2015comparing} found comparable performances with another state-of-the-art classical algorithm, Parallel Tempering \cite{hukushima1996exchange}. Due to its inherently parallel nature, PA is well-suited to be implemented on modern GPUs \cite{barash2017GPU}.
    \item \textbf{Global Annealing (GA)}~\cite{mcnaughton2020boosting, gabrie2022adaptive, ciarella2023machine} works similarly to Simulated Annealing, but instead of performing local, single-spin-flip moves, a generative model is used to propose global moves, in which all the spins are updated at the same time. These global moves are then accepted with a generalized version of the Metropolis criterion (see the Materials and Methods section for the details).\footnote{Note that, at variance with local MC moves, these global moves require the knowledge of the probability $\rho_\mathrm{NN}(\sigma)$ of the configuration $\sigma$ generated by the model. Therefore, in principle, only models for which $\rho_\mathrm{NN}(\sigma)$ can be computed (such as autoregressive models or normalizing flows) fit in this scheme.} The procedure starts with a population of configurations at high temperature, where sampling from \eqref{eq:GB} is easy. These configurations are used to train a generative model. The temperature is then lowered, and the previously trained network is used to propose moves at this lower temperature. After enough moves, the configurations are used to retrain the generative model. Then, the temperature is lowered once again, and the procedure continues. As in SA and PA, once the temperature is low enough, one considers the minimum energy configuration found as the estimated solution to \eqref{eq:argminH}. Notice that local MC moves can be alternated with the global ones to improve performances \cite{gabrie2022adaptive, del2025performance}. 
\end{itemize}

A schematic description of the three algorithms is given in Fig.~\ref{fig:algorithms}(a), which shows that they can be implemented in a very similar way, hence allowing for a fair comparison.
A more in-depth description is given in the Materials and Methods section for GA and in Supporting Information for SA and PA.

\new{We highlight the difference between Global Annealing and another widely used training scheme, Variational Neural Annealing \cite{hibat2021variational}. The latter approach aims at training the neural network by minimizing the \textit{reverse} Kullback–Leibler (KL) divergence between the GB distribution and the distribution obtained via the neural network. This approach does not require samples distributed according to the GB distribution and corresponds to minimizing the free energy of the network distribution. In practice, the network is first trained at high temperature (that is, minimizing the divergence with respect to GB at high $T$) and then progressively retrained at lower and lower temperatures. Instead, the GA method described above minimizes the \textit{direct} KL divergence, therefore requires data distributed according to the GB distribution. These data are obtained via a recursive series of Monte Carlo steps, as described above. While there is no direct comparison of the two procedures in the literature, the property of the direct KL to be \textit{mode covering} (with respect to the reverse KL, which is \textit{mode seeking})\footnote{\new{That is, the direct KL tends to be large for distributions that are zero when the target distribution is non-zero, while the reverse KL tends to be large for distributions that are non-zero where the target distribution is zero. So the former favors distributions that cover all the modes of the target, while the latter favors distributions that are only concentrated on (some of) the modes of the target, even without capturing all of them.}} suggests the GA procedure to be better suited for optimization problems.}

Fig.~\ref{fig:algorithms}(b) presents representative examples of the energy evolution with running time for the algorithms considered in this work. In this case, PA initially attains lower-energy approximate solutions, whereas only GA eventually reaches the exact \new{minimum energy configuration (MEC) (the details on how the MEC is found are reported in the Materials and Methods section)}. While this example provides a preliminary indication of a potential advantage of GA, it remains anecdotal; in the remainder, we examine this observation systematically through a detailed comparative analysis.

\begin{figure*}
    \centering
    \includegraphics[width=\linewidth]{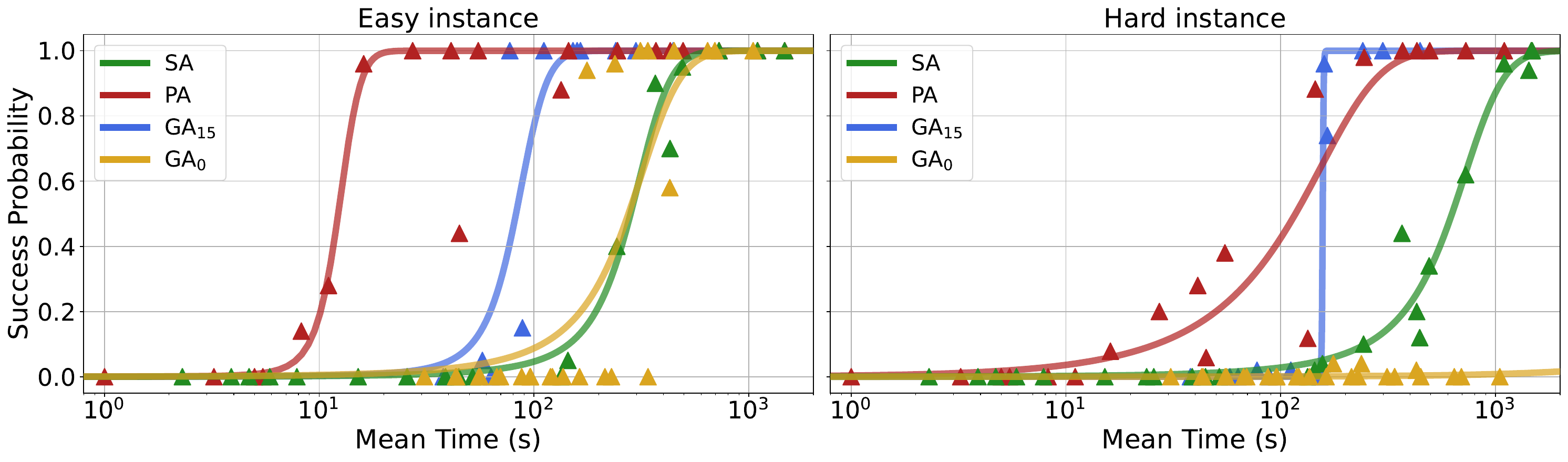}
    \caption{Success probability as a function of the mean running time for SA, PA and GA (with and without local moves), together with sigmoidal fits, for easy and hard realizations of the couplings at $N=10^3$. $\text{GA}_k$ identifies the GA algorithm with $k$ local MCSs per global move, so that $\text{GA}_0$ identifies the case in which no local moves are performed.}
    \label{fig:success_rates_twoinstances}
\end{figure*}

\section{Results}\label{sec:results}

We present a systematic comparison between Global Annealing (GA)---in which local moves alternate with global moves proposed by a generative model---and two state-of-the-art algorithms, Population Annealing (PA) and Simulated Annealing (SA). \new{The minimum energy is found either exactly for smaller system sizes or it is estimated using longer runs for larger system sizes (see the Materials and Methods section for the details).}
As described in the Algorithms section, in GA simulations, we employed the properly modified Metropolis acceptance criterion for global moves, which accounts for the probability that the generative model produces a given configuration (see Materials and Methods for details). We observed that neglecting this correction leads to a severe degradation in GA performance. This emphasizes the importance of using generative architectures for which the probability of generating a specific configuration can be efficiently computed.

\subsection{Local moves are essential}

We start by comparing GA with and without local moves. We call $\text{GA}_k$ the GA algorithm with $k$ local MCSs per global move (each sweep being a sequence of $N$ local MC moves), so that $\text{GA}_0$ identifies the case in which no local moves are performed.
We consider an easy and an hard\footnote{Here, \textit{easy} and \textit{hard} are loosely defined by simply looking at the time it takes for the different algorithms to find the ground state. Other criteria have been studied in the literature \cite{amey2018analysis}.} realization of the couplings $J_{ij}$ for a $d=3$ EA model on a cubic lattice of side $L=10$ with periodic boundary conditions, hence with $N = 10^3$ variables in total.

In Fig.~\ref{fig:success_rates_twoinstances} we plot the success probability as a function of the running time for $\text{GA}_0$ and $\text{GA}_{15}$ together with sigmoidal fits.  \textit{Easy} and \textit{hard} are here defined qualitatively, simply based on the time it takes for the algorithms to find the solution. 
The runs use a temperature schedule uniformly spaced in logarithmic scale.
The running time is varied by changing both the number of global moves and the number of temperatures, and the success probability is computed by running 50 independent runs and then checking which fraction reaches the minimum energy configuration, estimated as discussed in the Algorithms section.
We note that the same running time can correspond to different pairs of parameters (number of temperatures and number of local/global moves), which explains the non complete monotonicity of the scatter plot. Some longer runs correspond to choices of parameters that take a long time but are not effective at finding the minimum energy configuration.

The figure clearly shows that the addition of local moves is useful in finding the minimum energy configuration of the system.
Not including them leads to heavily deteriorated performances, with $\text{GA}_0$ almost always failing to find the minimum energy configuration in the hard case. 
We find that while $k=0$ is not a good choice, performance quickly saturates upon increasing $k$, hence $k=15$ is a suitable choice (see 
Supporting Information).

\subsection{Instance-to-instance fluctuations}

Having assessed the need for local moves, from now on we discard GA$_0$ and we focus on GA$_{15}$, which we call GA for simplicity.
In Fig.~\ref{fig:success_rates_twoinstances} we compare GA with SA and PA. We observe that GA outperforms SA on both instances, hence we will discard SA from subsequent discussions.
In contrast, while GA is outperformed by PA on the easy instance, the two methods achieve comparable performance on the hard instance.
This seems to suggest that, at this system size, PA performs better than GA on easy instances but on par or worse on harder instances.
Yet, we observe an important difference:
while PA starts having a non-zero probability of finding the minimum energy configuration before GA, it has a much less sharp transition from low to high success probability.  This is a first indication that GA is more robust than PA, in the sense that its outcome is more reproducible between distinct runs on the same instance (the success probability jumps very sharply from zero to one).
Additional results for other annealing schemes are reported as Supporting Information.

\begin{figure}[t]
    \centering
    \includegraphics[width=1\linewidth]{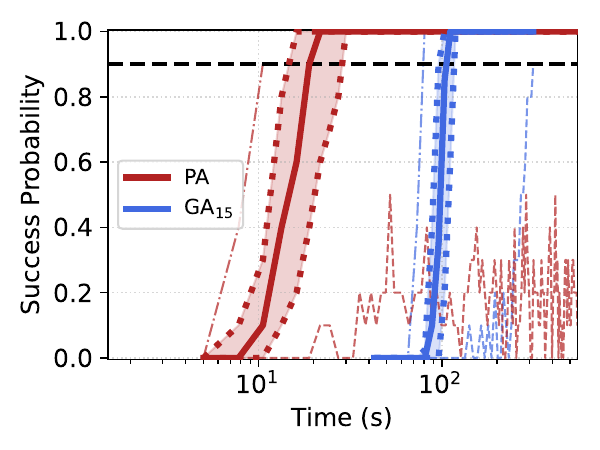}
    \caption{Median success probability (solid lines) over 200 instances for $N = 10^3$ instances as a function of the running time for PA (red) and GA (blue), together with the 25th and the 75th percentiles (dotted lines, shaded area) and the best (dash-dotted lines) and worst (color dashed) instances. The black dashed line corresponds to a 90\% success probability. To reduce computational cost, instances whose runs achieved a 90\% success rate were terminated, and a 100\% success rate was assumed thereafter for computing the average quantities. While PA tends to outperform GA on the majority of the runs, there are some instances, like the worst case shown here, in which PA performs much worse than GA. For PA, we have performed 10 MCS for every temperature, as in Ref.~\cite{wang2015comparing}. For GA, we performed 5 global moves per temperature, and 15 local MCS per global move. Both algorithms used a logarithmic spacing of the temperatures during annealing.}
    \label{fig:success_vs_time_L10}
\end{figure}

In order to obtain a more systematic assessment of the performance of GA versus PA, we consider an ensemble of 200 random independent realizations of the $J_{ij}$ and we repeat the analysis. As we have observed that increasing the number of temperatures improves the performance more than increasing the number of temperatures (especially for PA), here and in the following we vary the runtime by changing the number of temperatures in the annealing while keeping the other parameters fixed.
Because random instances of the EA model at the size we are now considering ($N=10^3$) are typically easy \cite{martinez2025problem}, PA tends to outperform GA on the majority of cases. This can be observed in Fig.~\ref{fig:success_vs_time_L10}, in which the median success rate over 200 realizations of the couplings $J_{ij}$, each estimated on 10 different runs, is plotted as a function of the runtime for PA and GA. Remarkably, as evidenced by the worst-case curves, GA tends to perform better on hard instances.  

\begin{figure}[t]
    \centering
    \includegraphics[width=1\linewidth]{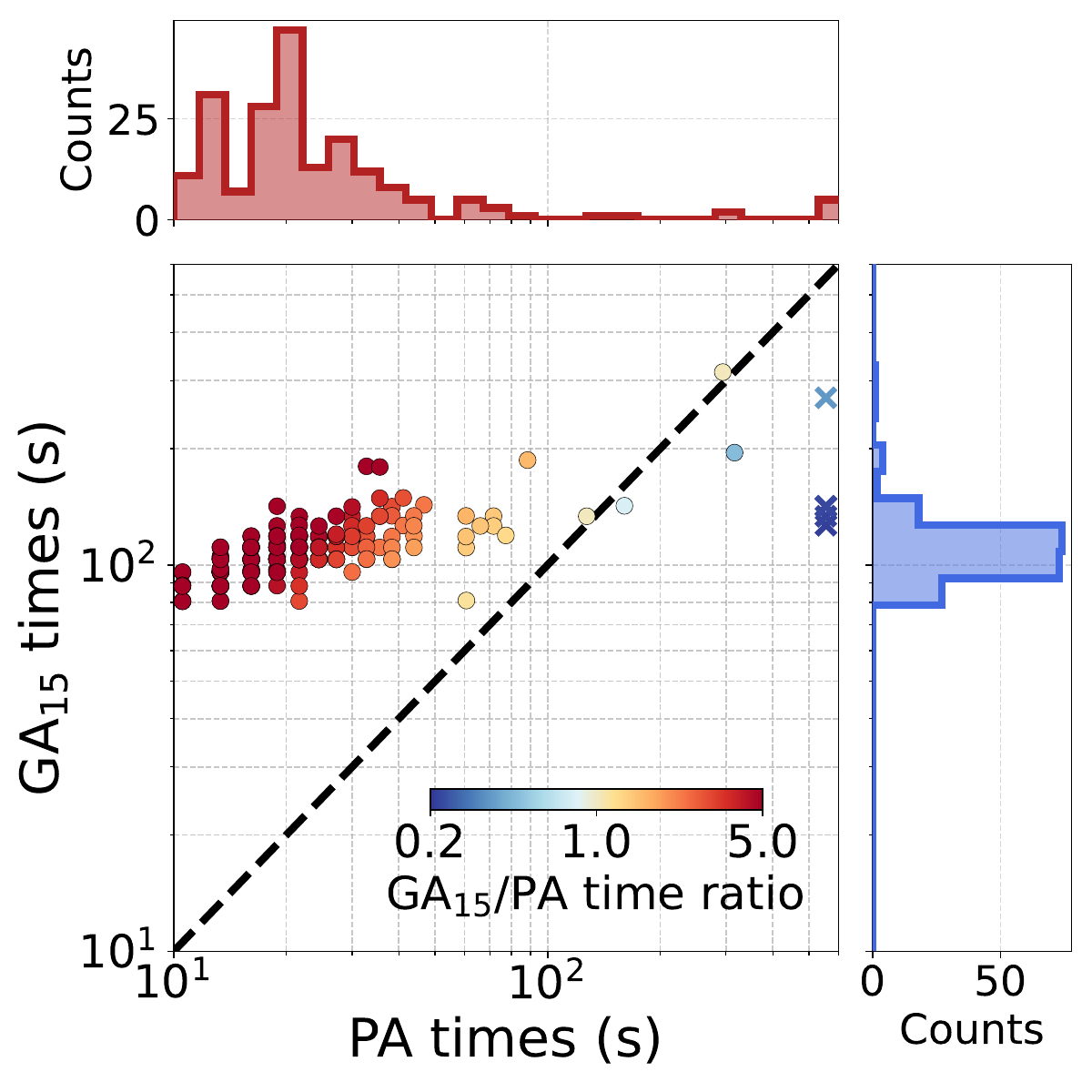}
    \caption{Scatter plot (in log-log scale) of the times it takes on each instance to reach a 90\% success rate with PA and $\text{GA}_{15}$, together with the corresponding histograms. Different colors of the datapoints highlight different ratios between the runtimes of the algorithms. The dashed line correspond to equal times. Crosses are points for which PA fails to achieve a 90\% success rate within the time limit of about 550 seconds. Data are the same as those used to obtain Fig.~\ref{fig:success_vs_time_L10}.
    }
    \label{fig:scatter_plot_times}
\end{figure}

This intuition is further confirmed by Fig.~\ref{fig:scatter_plot_times}, in which we compare the time it takes to reach a success rate equal or greater to 90\% for the different runs used in Fig.~\ref{fig:success_vs_time_L10}. In the majority of instances PA outperforms GA, mainly due to the cost of training the architecture. However, on some harder instances PA is outperformed by GA and in some cases even fails completely to reach a 90\% success rates within the given time limit (crosses in Fig.~\ref{fig:scatter_plot_times}). This highlights a greater robustness of the GA algorithm. Moreover, this result also shows the importance of including the training time of the generative model, which is often non-negligible, to achieve a fair comparison.

\subsection{Scaling to larger sizes}

\begin{figure}[t]
    \centering
    \includegraphics[width=1\linewidth]{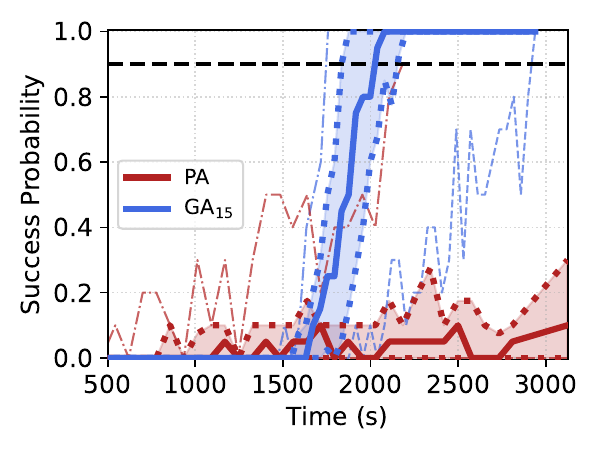}
    \caption{
    Same plot as in Fig.~\ref{fig:success_vs_time_L10}, here obtained with 10 instances with $N = 14^3$. The choices of the hyperparameters for both algorithms are the same as in Fig.~\ref{fig:success_vs_time_L10}.
    In this case, GA outperforms PA both in the median and in the best/worst case runs. 
    }
    \label{fig:success_vs_time_L14}
\end{figure}

The previous results seem to indicate a greater robustness of the GA algorithm with respect to PA when the problem hardness is increased, with the former not requiring tweaks in the hyperparameters. Yet, at $N=10^3$ PA remains more efficient in most cases. We thus
tested the two algorithms for larger (and therefore harder) instances with $L=14$, hence $N=14^3=2744$. 
In Fig.~\ref{fig:success_vs_time_L14} we consider the success rate as a function of running time for 10 different such instances. We clearly see that in this case GA outperform PA when the same hyperparameters of the $N = 10^3$ case are used. While it is known that for PA one has to increase the population size when $N$ increases~\cite{machta2010pa}, this result confirms the better robustness of GA to changes in the problem specification. Moreover, additional tests on PA with a half-as-big or three-times-bigger population did not seem to yield better performances when the total runtime is taken into account.

Fig.~\ref{fig:success_vs_time_L14} presents clear evidence that an algorithm exploiting machine-learning techniques can be much more effective than state-of-the-art classical algorithms. The difference in performance between PA and GA is remarkable, with the worst run of GA taking less than 3000 seconds, and being not far from the best run of PA that takes more than 2000 seconds.

\subsection{Overlap probability distribution}

\begin{figure*}
    \centering
\includegraphics[width=.9\linewidth]{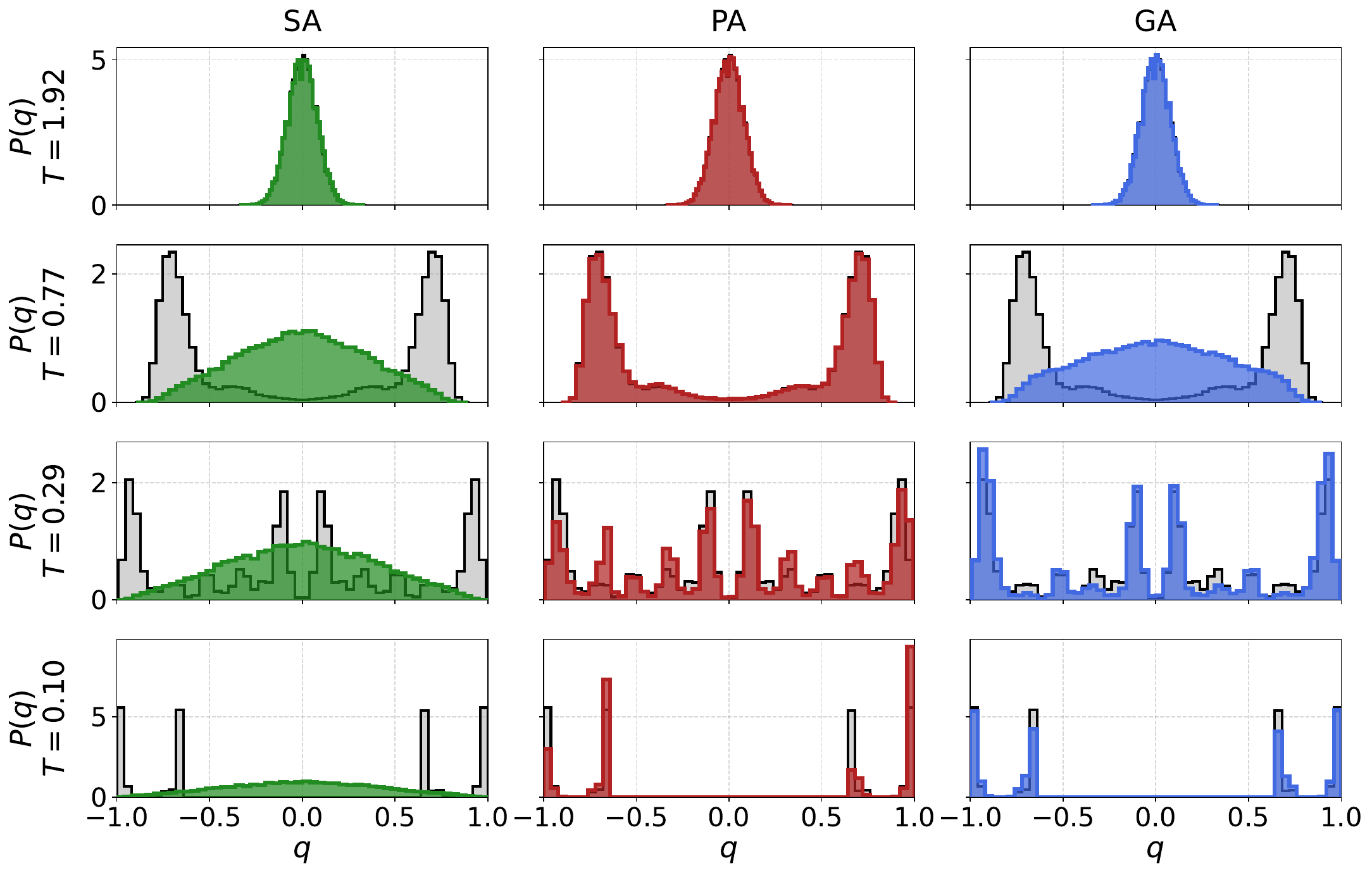}
    \caption{Probability density of the overlap as obtained by SA, PA, GA (green, red and blue, respectively) compared to the one obtained by a much longer PA run (gray) at different temperatures for different runs at which the energy minimum is reached. The three different algorithms use runs of different lengths, chosen in order to correspond approximately to the first time at which a 100\% success rate is achieved.}
    \label{fig:Pq}
\end{figure*}

It is interesting to understand the mechanism that allows SA, PA and GA to find the minimum energy. To this aim, for each algorithm we consider one successful run on the hard instance 
of Fig.~\ref{fig:success_rates_twoinstances}. At each temperature, we consider the set of $M=2^{17}$ configurations that are annealed in parallel by each algorithm.
The overlap (similarity) between two such configurations $\sigma^1$ and $\sigma^2$ is defined as
\begin{equation}
    q = \frac{1}{N}{\sigma}^1 \cdot {\sigma}^2 = \frac{1}{N} \sum_{i = 1}^N \sigma^1_i \sigma^2_i \ ,
\end{equation}
and is a key quantity to verify whether the system has reached thermal equilibrium, i.e. if the distribution over the configurations is given by ~\eqref{eq:GB}.
In Fig.~\ref{fig:Pq} we compare the probability density of the overlap for SA, PA, and GA during annealing with the equilibrium distribution obtained from a much longer run of PA. 

Interestingly, we observe that SA finds the minimum but remains out of equilibrium at all temperatures, producing distributions that differ significantly from the reference equilibrium ones. PA follows the correct distribution more closely across most temperatures, but at the lowest temperatures it fails to reproduce the relative weights of the peaks and even breaks the expected symmetry around $q = 0$. GA shows the opposite trend: it struggles to match the distribution at intermediate temperatures (mainly due to the large temperature steps and the few training epochs), but at low temperatures it captures the peak weights much more accurately and preserves the symmetry, yielding a closer agreement with the equilibrium reference.

These different mechanisms highlight the distinct ways in which the minimum energy configuration is reached. In SA, the different elements of the population have no means of communicating with one another. As a result, the ensemble as a whole cannot thermalize efficiently. Finding the minimum then becomes essentially a matter of rare events: by chance, one or a few elements of the population may end up in the correct configuration. In contrast, both PA and GA include mechanisms that allow information to be exchanged among different elements of the population. In PA this occurs through the reweighting step, in which lower-energy configurations are preferentially replicated. In this way, population elements that discover good states effectively share this information with others, leading to improved thermalization and facilitating the discovery of lower-energy configurations.
However, this requires an increasing population size for larger instances, to maintain sufficient diversity. 
In GA this information-sharing role is played by the generative model, which extracts information from the entire population and can then propose new moves accordingly.

\section{Discussion}\label{sec:discussion}

In this work, we studied the application of the machine-learning-assisted Global Annealing (also previously called Sequential Tempering) to the optimization of spin glass systems in finite dimension, which provide hard instances in the QUBO class. Specifically, we tested the capabilities of the algorithm on the NP-hard problem of finding the minimum energy configurations of the Edwards-Anderson model in three dimensions. We considered system sizes of $N = 10^3$ and $N = 14^3$, comparable or larger than state-of-the-art studies in the field.

First, we verified the theoretical prediction \cite{del2025performance} that standard local Metropolis moves are needed to improve the performance of Global Annealing. An intuitive explanation is as follows. If the generative model proposes moves with the correct GB probability, $\rho_{\text{NN}} \propto e^{-\beta H(\sigma)} $, then global moves become equivalent to temperature-swap moves in Parallel Tempering (PT)~\cite{hukushima1996exchange}: indeed, the acceptance probability described in ~\eqref{eq:prob} of the Materials and Methods section reduces to the temperature-swap acceptance probability of PT. This establishes a direct analogy between GA and PT: in GA, global moves take the role of temperature swaps, while the entire ladder of replicas in temperature is effectively replaced by a single neural network (a substitution that underlies the potential speedup of the GA approach). A key feature of PT is that temperature-swap moves are alternated with local updates. By analogy, this provides a clear motivation for why local moves remain essential in the GA procedure. Additionally, this parallelism hints at the reason why GA is effective: it substitutes the long (and computationally expensive to run) ladder of temperatures of PT with a single generative model.

Moreover, we compared the GA technique with two classical algorithms, Simulated Annealing and Population Annealing. Within a comparable \texttt{torch} implementation, we found that GA consistently outperforms SA.
Instead, the comparison between GA and PA is less obvious. While we observed that PA outperforms GA on the majority of the $N = 10^3$ instances, we also noticed that GA is extremely robust and its running time depends weakly on the hardness of the instance.
Finally, for some hard $N = 10^3$ instances and for all of the $N = 14^3$ instances, GA shows a clear superiority over PA, requiring much shorter wall-clock times to reach the energy minimum.
As noted in the Introduction, demonstrating the advantage of machine-learning methods over classical algorithms in hard optimization problems has been a long-standing goal. Our results provide clear evidence of this advantage.

\new{While the main focus of this work is to compare GA with the state-of-the-art annealing algorithms, it would be interesting to characterize the scaling of resources required by this class of annealing-based algorithms to find the exact minimum energy. This has recently been attempted in optimization problems where the minimum energy is known \cite{angelini2026algorithmic}, but in the general case, much more work is needed, and we leave this open question for future work.}

We stress that all algorithms have been implemented using the \texttt{python} library \texttt{torch} \cite{NEURIPS2019_9015} and run on a single GPU with the same level of optimization, to make the comparison fair.
Nonetheless, the algorithms' implementations could be further improved. For instance, SA and PA could benefit from implementations in Cuda C \cite{cuda-c-programming-guide} and multi-spin coding \cite{jacobs1981multi}, while GA could become faster by the use of \texttt{jax} \cite{jax2018github} or \texttt{torch.compile} to achieve a faster training of the generative model.
The GA implementation in this work uses the shallow MADE architecture described in the Materials and Methods section. This architecture has a number of parameters scaling as $N^2 = L^6$ in the $d=3$ case. The use of lighter architectures, such as TwoBo~\cite{biazzo2024sparse}, the three-dimensional HAN~\cite{bialas2025hierarchical} or 4N~\cite{del2025nearest}, could be taken into account to achieve a better scaling.
\new{Future experiments could also involve more powerful autoregressive networks, such as transformers \cite{vaswani2017attention} or MAMBA \cite{gu2024mamba}. While these architectures are probably more expressive and could better fit the target distribution, it is not obvious that they would achieve better computational performance: indeed, their training could require orders of magnitude more time (e.g., for the transformer architecture introduced in~\cite{bunaiyan2025isingformer}), thus severely degrading the overall performance. More advanced architectures could also be developed to incorporate the disorder realization as an additional input, as done, for example, in Ref.~\cite{viteritti2025quantum} for the quantum setting. Such architectures would have the advantage of not requiring retraining for each disorder instance, which could in turn improve performance at inference time. Whether these approaches are feasible, applicable to annealing schemes, and more efficient than the one adopted here, especially when one scales the system to progressively larger sizes, remains an open question and a subject for future work.}

In this study, we have used GA only for the task of finding the minimum energy. However, GA is much more general, and can also be used for sampling from \eqref{eq:GB} at any given temperature. The question of whether GA is effective in sampling is still open~\cite{mcnaughton2020boosting, ciarella2023machine}. The results shown in Fig.~\ref{fig:Pq} suggest that GA can find the minimum energy configurations even if the procedure does not sample equilibrium configurations at the intermediate temperatures. The equilibration (or mixing) times of GA should be more systematically compared to state-of-the-art algorithms.

Finally, in this study, we have considered only one ML-assisted algorithm, Global Annealing, and two classical algorithms, Simulated Annealing and Population Annealing. Many additional algorithms can be tested and compared, as described in the Introduction. Future work should focus on constructing proper benchmarks for a systematic comparison of existing algorithms, to avoid unsubstantiated claims of superiority.

\section{Materials and Methods}

\subsection{General details}

All annealing procedures begin at a high temperature of $T = 1.92$, which is gradually reduced to a low temperature of $T = 0.1$ following a logarithmically spaced schedule.

Local MC updates are carried out in a checkerboard scheme, where spins with odd and even indices are alternately updated in parallel. In this way, a MCS is made of a single odd-indices move followed by a single even-indices move, the two moves combined proposing a flip for all the $N$ spins.

Each algorithm is run with a population of $2^{17} = 131072$ configurations and is implemented in \texttt{torch}~\cite{NEURIPS2019_9015}. The initial configurations at $T = 1.92$ are assumed to be provided, and they are thermalized through 200 MCS. The runtime of this initial thermalization step is excluded from the reported timings of all algorithms.

Reported runtimes refer to runs on single NVIDIA Tesla V100-SXM2-32GB GPUs for all data in the articles except for the example data in Fig.~\ref{fig:algorithms}, which were obtained on a NVIDIA Tesla V100S-PCIE-32GB.

\subsection{General description of the Global Annealing procedure}

In the GA procedure, one uses a generative model to generate configurations $\sigma'$ of the system, approximately at equilibrium, i.e., according to \eqref{eq:GB} at a temperature $\beta$. These configurations are then used as global proposal moves for the MC procedure at $\beta' =  \beta + \new{\Delta \beta (\beta)} > \beta$, instead of the standard single-spin-flip moves of the local MC algorithm. The move $\sigma \to \sigma'$ is then accepted with an acceptance probability:
\begin{equation}\label{eq:prob}
\begin{split}
    \text{Acc}\left[{\sigma} \rightarrow {\sigma}'\right] &= \min\left[1, \frac{\rho_\text{GB}({\sigma}') \times \rho_{\text{NN}}({\sigma})}{\rho_\text{GB}({\sigma}) \times \rho_{\text{NN}}({\sigma}')}\right], 
    \end{split}
\end{equation}
where $\rho_\text{NN}$ is the probability that the generative model generates a configuration $\sigma$. This choice for the acceptance rate guarantees, under ergodicity assumptions, that the distribution over the states is asymptotically given by \eqref{eq:GB}. The advantage of proposing global moves with the generative model is that all spins are updated simultaneously, making the procedure, in principle, much faster at sampling independent configurations than when only local moves are used. Notice that, if one neglects the ratio $\rho_{\text{NN}}({\sigma})/\rho_{\text{NN}}({\sigma'})$ in \eqref{eq:prob}, detailed balance is no longer satisfied, hence the convergence to \eqref{eq:GB} is not guaranteed anymore. Correspondingly, we found that the procedure's performance worsens substantially (not shown).

The general scheme of the Global Annealing procedure is sketched in Fig.~\ref{fig:algorithms}. It is summarized in Alg.~\ref{alg:GA}.

\begin{algorithm}
\caption{Global Annealing}
\label{alg:GA}
\begin{algorithmic}[1]
    \State \textbf{Input:} Initial inverse temperature $\beta_{\text{start}}$, final inverse temperature $\beta_{\text{end}}$, temperature \new{schedule} $\new{\Delta \beta (\beta)}$, number of configurations~$M$, number of global steps per temperature $\theta_g$, number of local steps per global step $\theta_l$.
    \State \textbf{Initialize:} A set of $M$ equilibrium configurations at $\beta_{\text{start}}$ (sampled e.g. using standard Metropolis MC)
    \While{$\beta < \beta_{\text{end}}$}
        \State Train a neural network (NN) using the set of $M$ configurations
        \State Lower the temperature: $\beta \gets \beta + \new{\Delta \beta (\beta)}$
        \For{$m$ in $1,\dots,M$}
        \State Choose the $m$-th configuration from the set as the initial state
        \For{$t$ in $1, \dots, \theta_g$}
            \State Propose a new configuration using the NN
            \State Accept or reject the configuration with probability \eqref{eq:prob} at the new temperature $T=1/\beta$
            \For{$t$ in $1, \dots, \theta_l$}
            \State Perform a local MC step
        \EndFor
        \EndFor
        \EndFor
    \EndWhile
\end{algorithmic}
\end{algorithm}

\subsection{Details on the architecture}

In this work, we have used a shallow MADE (Masked Autoencoder for Distribution Estimation, \cite{germain2015made}) autoregressive architecture, which is modeled as:
\begin{equation}
    P(\sigma_i | \sigma_{<i}) = \frac{\exp\left( \sum_{j=1}^{i-1} W_{ij} \sigma_i \sigma_j \right)}{2 \cosh\left( \sum_{j=1}^{i-1} W_{ij} \sigma_j \right)},
\end{equation}
where $\sigma_{<i}$ is the set of spins $\sigma_{<i} = \{\sigma_1, \dots, \sigma_{i-1} \}$. We note that the autoregressive approach requires choosing an ordering of the variables. Here, spins are taken in raster order, i.e., from left to right, line by line, and plane by plane.

In practice, the architecture consists of a dense autoregressive layer followed by a sigmoidal activation and has $\mathcal{O}(N^2)$ parameters. This relatively simple design allows us to focus on the annealing procedure rather than on the architectural details. As mentioned in the Discussion section, alternative choices of the generative architecture could further improve the GA procedure. However, it is not obvious that more sophisticated architectures would necessarily yield better results, since (depending on the type of the architecture)  their training could require orders of magnitude more time (as the transformers architecture introduced in \cite{bunaiyan2025isingformer}), which would severely degrade overall performance.

\subsection{Details on the training procedure}

Training is performed by minimizing the binary cross-entropy loss (i.e., minimizing the KL divergence between $\rho_\text{NN}$ and $\rho_\text{GB}$). The initial training runs for $40$ epochs using the Adam optimizer with learning rate $\eta_0 = 10^{-3}$. We employ an exponential learning–rate schedule that halves the rate every $10$ epochs. Early stopping is applied on the training set solely as a plateau detector for the training objective, with a patience of 10 epochs. Each epoch processes the full set of $2^{17}$ configurations in mini-batches of size $256$. For retraining at lower temperatures, we perform a single epoch per stage without the previously mentioned regularizations.

\subsection{\new{Best estimate of the minimum energy configuration}}

\new{We consider instances of the $d=3$ EA model with Gaussian couplings $J_{ij}$, for which the solution to \eqref{eq:argminH}, i.e., the minimum energy configuration (MEC) or  \textit{ground state}, is unique with high probability.
Finding the MEC may require a time scaling super-polynomially in $N$ (as the problem is NP-hard), but we need a fast and reliable estimate of the MEC to perform the comparison among algorithms.
For each instance, we define the MEC ``best estimate'' as the lowest energy state found by \emph{all} the runs of GA, SA, and PA that we have performed.
In many cases, we are confident these algorithms found the exact MEC, while in other cases, they may have found only a very good approximation to the MEC.
In any case, the comparison among algorithms is fair and meaningful.}

\new{In particular, for the $N = 10^3$ instances shown in Figs.~\ref{fig:success_vs_time_L10}-\ref{fig:scatter_plot_times}, the MEC has been estimated by performing a set of ten longer runs of GA, with 10 global steps, 30 local steps per global step, and roughly 35 temperatures, each run taking approximately 800 seconds.
For the $N = 14^3$ instances shown in Fig.~\ref{fig:success_vs_time_L14}, about 60 temperatures have been used, and each run took about 7000 seconds.}

\new{For Fig.~\ref{fig:success_rates_twoinstances} and for many of the instances in Figs.~\ref{fig:success_vs_time_L10}-\ref{fig:scatter_plot_times} we checked that the MEC found with the above criterion coincides with the one obtained by the Gurobi solver~\cite{gurobi}, run with a 0\% gap setting, which guarantees convergence to the exact MEC at the price of an unfavorable scaling of the running time (typically exponential in the problem size).
As a further test, in the $L = 10$ case, we solved 2500 more instances using the same hyperparameters described above (one run per instance).
The mean ground state energy found across samples, $-1.6977(5)$, is in excellent agreement with the value $-1.6980(3)$ reported in Ref.~\cite{wang2015comparing}.}

\new{For the above reasons, and since for long enough annealing times the best-performing algorithms consistently find the same configuration, we are confident that our best estimate for the MEC is often the exact ground state. We cannot exclude that, in some cases, another lower energy configuration exists, which is never found by any of the algorithms we tested. However, even in these cases, where our best estimate is just a very low energy configuration, the comparison among algorithms is meaningful and fair (which is the aim of the present work).}

\section{Data Availability}
The code and data used in this paper are available at the GitHub repository \url{https://github.com/Laplaxe/MLMC_optimization}.

\begin{acknowledgments}

We warmly thank Stefano Bae, Indaco Biazzo, Giulio Biroli, Giuseppe Carleo, Patrick Charbonneau, Marylou Gabri\'e, and Enzo Marinari, for many interesting discussions related to this work. The research has received financial support from the “National Centre for HPC, Big Data and Quantum Computing”, Project CN\_00000013, CUP B83C22002940006, NRRP Mission 4 Component 2 Investment 1.4,  Funded by the European Union - NextGenerationEU. LMDB acknowledges funding from the Bando Ricerca Scientifica 2024 - \textit{Avvio alla Ricerca} (D.R. No. 1179/2024) of Sapienza Università di Roma, project B83C24005280001 – MaLeDiSSi. We acknowledge support from the computational infrastructure DARIAH.IT, PON Project code PIR01\_00022, National Research Council of Italy.

\end{acknowledgments}

\bibliographystyle{apsrev4-2}
\bibliography{refs}

\appendix

\onecolumngrid

\renewcommand{\thesection}{S\arabic{section}} 
\renewcommand{\appendixname}{}
\renewcommand{\thefigure}{S\arabic{figure}}
\setcounter{figure}{0}

\newpage

\centerline{\large\bf Supplementary Information}

\section{Additional numerical results for the $L = 10$ easy/hard case}

\begin{figure}[b]
    \centering
    \includegraphics[width=0.8\linewidth]{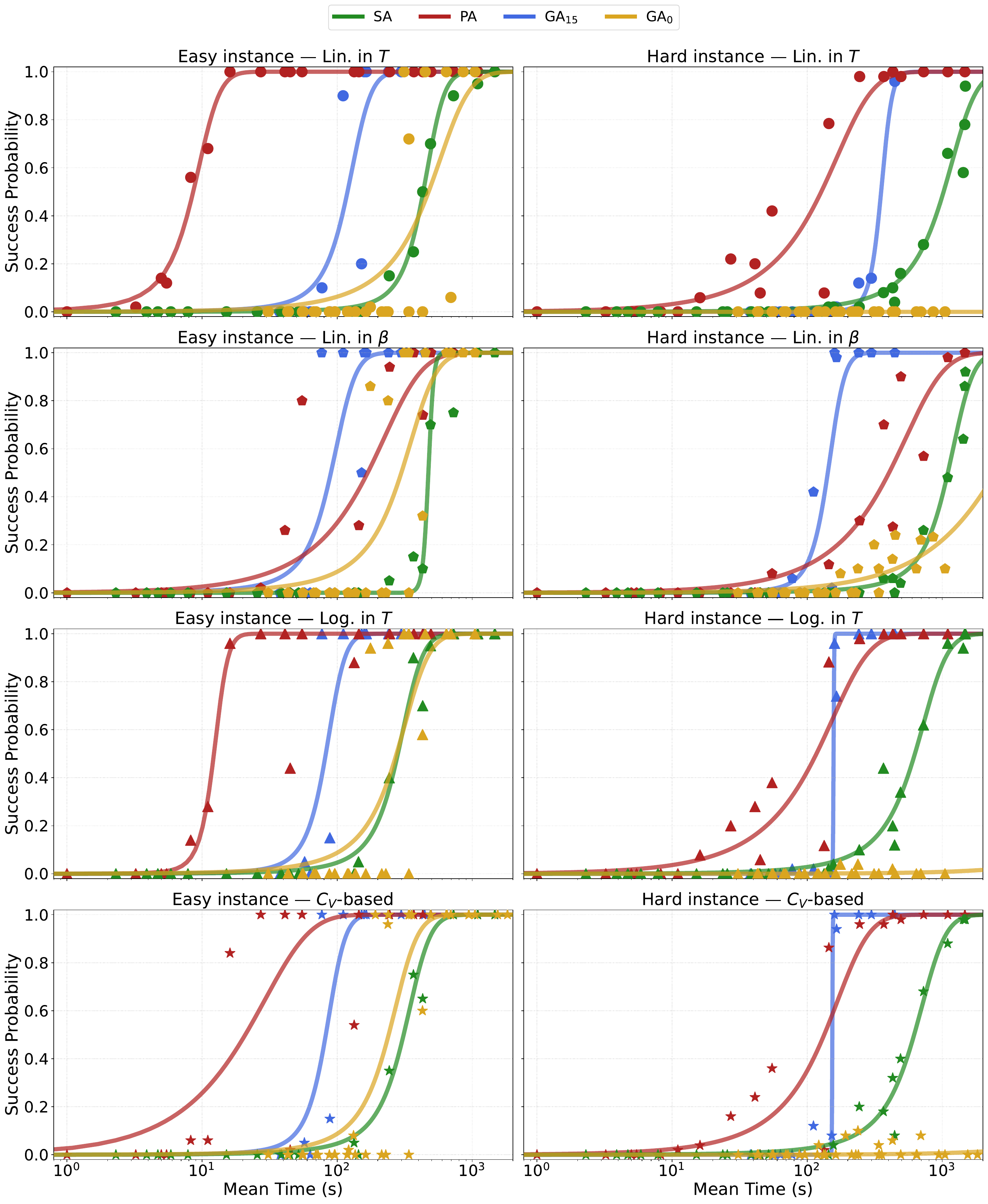}
    \caption{Same results as in Fig.~2 of the main text, but considering more temperature schedules for both the easy instance (left column) and the hard instance (right column). The temperature schedules are (from top to bottom) linear in temperature ($\Delta T = \text{const} \implies \Delta \beta \propto \beta^2$), linear in the inverse temperature ($\Delta \beta = \text{const}$), logarithmic in temperature ($\Delta\beta \propto \beta$) and based on the specific heat of the system ($\Delta \beta \propto C_V^{-1/2}$).}
    \label{fig:extensive_results_twoinstances}
\end{figure}

In Fig.~\ref{fig:extensive_results_twoinstances} we show the same plots of Fig.~2 of the main text, but considering more temperature schedules in addition to the logarithmically spaced one: namely, linear in temperature, linear in the inverse temperatures, and based on the specific heat of the system (as detailed in the caption). 
To approximate the heat capacity of the model, we use the following empirical expression that fits well the numerical data at equilibrium
\begin{equation}
C_V(T,N) \;=\; N \,\frac{34.19\, T^{2}}{\bigl(10.36 + T^{3}\bigr)^{2}}\,.
\end{equation}
The $C_V$-based temperature schedule follows from the well–known criterion for parallel tempering that neighboring replicas should be spaced in temperature according to $\Delta \beta \propto 1/\sqrt{C_V}$ to ensure a roughly uniform swap rate \cite{pelissetto2014large}. Note that this choice satisfies the requirement $\Delta \beta \propto 1/\sqrt{N}$, which is known to be a good temperature schedule for Global Annealing \cite{del2025performance}.
\new{For completeness, we point out that even more advanced schedules, such as the one presented in \cite{amey2018analysis}, can be used (but we do not expect our result to change)}.

\begin{figure}[t]
    \centering
    \includegraphics[width=0.5\linewidth]{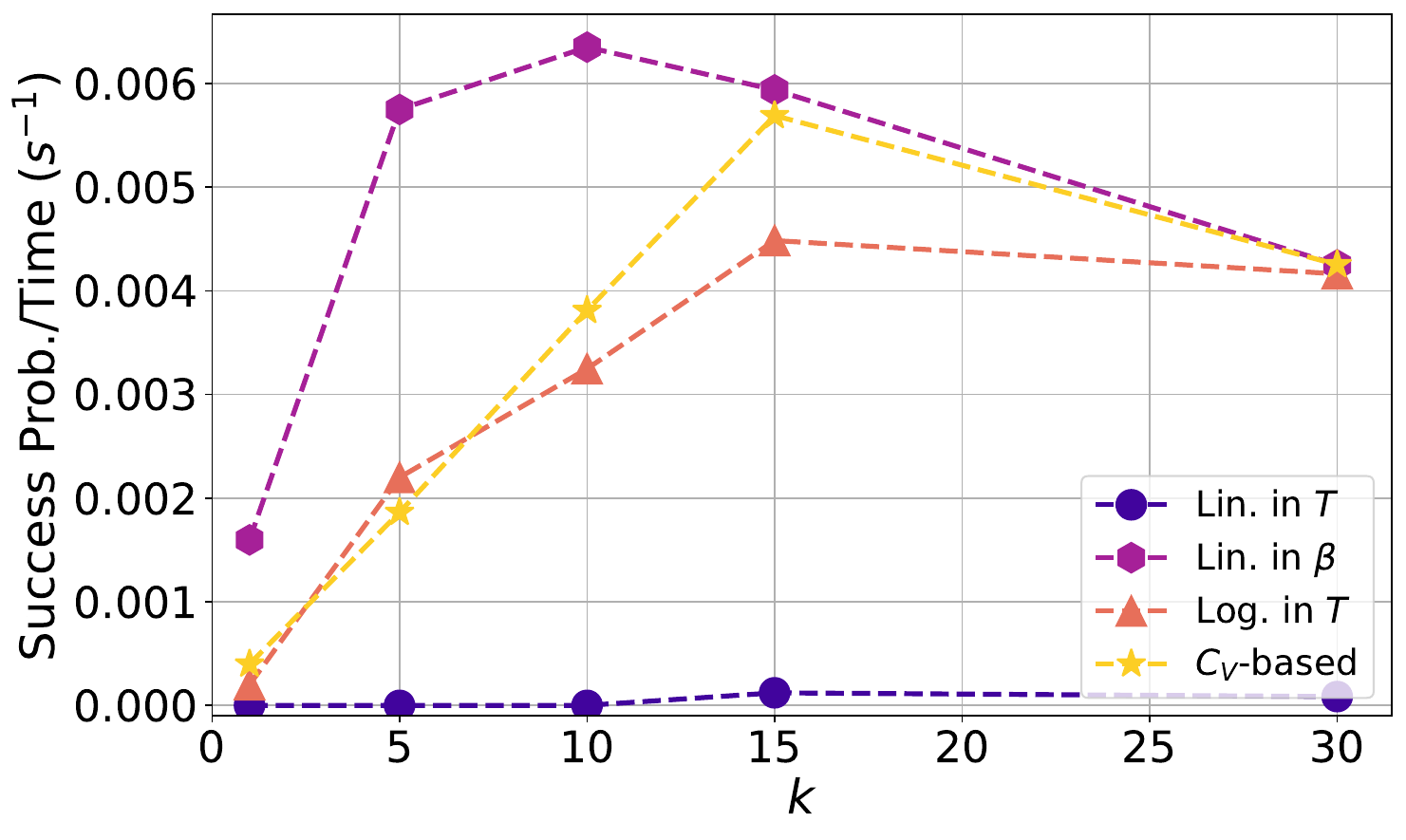}
    \caption{Success probability over runtime as a function of the number of local steps per local steps $k$ for the GA algorithm.}
    \label{fig:best_k}
\end{figure}

Additionally, in Fig.~\ref{fig:best_k} we plot the success probability divided by runtime for different values of the number $k$ of local steps per global step in the $\text{GA}_k$ algorithm. The optimal $k$ value for the logarithmic schedule is 15, which is the one used in the main text for the comparisons between GA and PA.

\new{The reader may be surprised to notice that other schedules (linear in $\beta$ and $C_V$-based) have better performances. We opted for the logarithmic schedule as this turns out to be the best schedule for PA and we do not want to give any unfair advantage to GA. However, when using GA in pratice, we advice to opt for the optimal scheduling.}

\section{Classical algorithms}

\subsection{Local Markov Chain Monte Carlo}

In the standard local Metropolis Monte Carlo \cite{newman1999monte} algorithm, one performs a series of local, single-spin-flip updates. A single Monte Carlo step consists of the following operations, starting from a configuration $\sigma(t) = \sigma$ at time $t$:
\begin{enumerate}
    \item propose a new configuration by flipping the spin at a randomly chosen site, $\sigma_i \rightarrow -\sigma_i$;
    \item calculate the energy difference 
    \[
    \Delta E = 2 \sigma_i \sum_{j \in \partial i} J_{ij}\,\sigma_j ,
    \]
    where the sum runs over the nearest neighbors $\partial_i$ of site $i$ and $J_{ij}$ are the quenched couplings of the Edwards--Anderson model;
    \item accept the move, i.e. set $\sigma \gets\sigma'$, where $\sigma'$ is obtained from $\sigma$ by flipping $\sigma_i$, with probability
    \[
    \text{Acc}\left[\sigma \rightarrow \sigma'\right] = \min\left[1, e^{-\beta \Delta E}\right] \, ;
    \]
    otherwise, reject the move and set $\sigma(t+1)=\sigma$.
\end{enumerate}
$N$ such steps are commonly referred to as a Monte Carlo sweep (MCS). The computational complexity of a MCS is $\mathcal{O}(N)$. The previously described local MC moves can be performed at a fixed temperature $\beta$. In this case, the distribution over all configurations will asymptotically match the correct Gibbs-Boltzmann one, $\rho_\text{GB}$. However, this procedure can in practice be very slow, especially at low temperature. One way to circumvent the problem is to use some kind of annealing procedure, in which one starts at high temperature and then progressively lowers it.

\subsection{Simulated Annealing}

In simulated annealing, the spins are updated using the local moves described in the previous paragraph. The temperature, however, is not fixed, but is progressively lowered according to a chosen schedule. As $\beta$ increases, the system tends to reach configurations of lower energy, providing increasingly accurate approximations to the minimum energy configuration.

While the original formulation of simulated annealing evolves only a single configuration of the system at the time, we consider a straightforward generalization in which a population of $M$ configurations is evolved in parallel, in order to make the comparison with PA and GA more fair.

The general scheme of the Simulated Annealing procedure, which is sketched in Fig. 1 of the main text, is summarized in Alg. \ref{alg:SA}.

\begin{algorithm}
\caption{Simulated Annealing}
\label{alg:SA}
\begin{algorithmic}[1]
    \State \textbf{Input:} Initial inverse temperature $\beta_\text{start}$, final inverse temperature $\beta_\text{end}$, temperature step $\Delta\beta$, number of configurations~$M$, number of MCS for temperature $\theta_\text{l}$.
    \State \textbf{Initialize:} A set of $M$ equilibrium configurations at $\beta_\text{start}$ (sampled e.g. using standard Metropolis MC)
    \While{$\beta < \beta_\text{end}$}
        \State Lower the temperature: $\beta \gets \beta + \Delta \beta$
        \For{$t=1$ \textbf{to} $\theta_l$}
            \State Perform a local MC step
        \EndFor
    \EndWhile
\end{algorithmic}
\end{algorithm}

\subsection{Population Annealing}

While the original formulation of simulated annealing employs only a single configuration, we consider a straightforward generalization in which a population of $M$ configurations is evolved in parallel. 

Population Annealing (PA) follows the same general idea of progressively lowering the temperature, but it introduces a reweighting step at each temperature. When the inverse temperature is updated from $\beta$ to $\beta'$, each configuration $i$ in the population, with energy $E_i$, is assigned a weight
\[
w_i = \exp\!\big[-(\beta' - \beta)\,E_i\big] .
\]
These weights determine the expected number of copies of each configuration in the next population: low-energy configurations are preferentially replicated, allowing information about good states to propagate across the population, while high-energy configurations are gradually eliminated. 

The general scheme of the PA procedure is sketched in Fig. 1 of the main text and summarized in Alg.~\ref{alg:PA}.
There is some freedom of the choice for the resampling procedure \cite{gessert2023resampling}. In this work we considered multinomial resampling as is straightforwardly implemented in \texttt{torch} via the \texttt{torch.multinomial} function.

\begin{algorithm}
\caption{Population Annealing}
\label{alg:PA}
\begin{algorithmic}[1]
    \State \textbf{Input:} Initial inverse temperature $\beta_\text{start}$, final inverse temperature $\beta_\text{end}$, temperature step $\Delta\beta$, number of configurations $M$, number of MCS per temperature $\theta_l$.
    \State \textbf{Initialize:} Set $\beta \gets \beta_\text{start}$. Prepare $M$ equilibrium configurations at $\beta$
    \While{$\beta < \beta_\text{end}$}
         $\beta' \gets \beta + \Delta\beta,\, $
        \State compute the energies of the $M$ configurations, $\{E_i\}_{i=1}^M$
        \State For each configuration $i$, set $w_i \gets \exp\!\big(-(\beta' - \beta)\,E_i\big)$
        \State Normalize the weights $\tilde{w}_i \gets w_i/\sum_{i=1}^M w_i$
        \State Resample the $M$ configurations according to the set of weights $\{\tilde{w}_i\}_{i=1}^{M}$
        \For{$\ell=1$ \textbf{to} $\theta_l$}
            \State Perform a local MC step on the $M$ configurations (in parallel)
        \EndFor
        \State Set $\beta \gets \beta'$
    \EndWhile
\end{algorithmic}
\end{algorithm}

\end{document}